\documentclass[pra,aps,twocolumn,showpacs]{revtex4-2} 
\usepackage{amsmath,amssymb,amsthm}
\usepackage{graphicx}
\usepackage{subfigure}
\usepackage{amsmath}
\usepackage{amsfonts,amssymb}
\usepackage{bm}
\usepackage{float}
\usepackage{color}
\usepackage{sidecap}
\allowdisplaybreaks
\usepackage{soul}
\setstcolor{red}
\usepackage[normalem]{ulem}
\usepackage{hyperref}
\hypersetup{colorlinks=true,breaklinks,urlcolor=blue,linkcolor=blue,citecolor=blue}

\begin{document}

\title{Gaussian eigenstate pinning in non-Hermitian quantum mechanics}
\author{Qi-Bo Zeng}
\email{zengqibo@cnu.edu.cn}
\affiliation{Department of Physics, Capital Normal University, Beijing 100048, China}

\author{Rong L\"u}
\affiliation{State Key Laboratory of Low-Dimensional Quantum Physics, Department of Physics, Tsinghua University, Beijing 100084, China}
\affiliation{Frontier Science Center for Quantum Information, Beijing 100084, China}

\begin{abstract}
We study a one-dimensional system subjected to a linearly varying imaginary vector potential, which is described by the single-particle continuous Schr\"odinger equation and is analytically solved. The eigenenergy spectrum is found to be real under open boundary condition (OBC) but forms a parabola in the complex energy plane under periodic boundary condition (PBC). The eigenstates always exhibit a modulated Gaussian distribution and are all pinned on the same position, which is determined by the imaginary vector potential and boundary conditions. These behaviors are in sharp contrast to the non-Hermitian skin effect (NHSE) in systems with constant imaginary vector potential, where the eigenstates are exponentially distributed under OBC but become extended under PBC. We further demonstrate that even though the spectrum under PBC is an open curve, the Gaussian type of NHSE still has a topological origin and is characterized by a nonvanishing winding number in the PBC spectrum. The energies interior to the parabola can support localized edge states under semi-infinite boundary condition. The corresponding tight-binding lattice models also show similar properties, except that the PBC spectrum forms closed loops. Our work opens a door for the study of quantum systems with spatially varying imaginary vector potentials.  
\end{abstract}
\maketitle
\date{today}

\section{Introduction}
Over the past few decades, a significant amount of research has been devoted to non-Hermitian Hamiltonians~\cite{Cao2015RMP,Konotop2016RMP,Ganainy2018NatPhy,Ashida2020AiP,Bergholtz2021RMP}, which can effectively describe open systems in both  classical~\cite{Makris2008PRL,Klaiman2008PRL,Guo2009PRL,Ruter2010NatPhys,Lin2011PRL,Regensburger2012Nat,Feng2013NatMat,Peng2014NatPhys,Wiersig2014PRL,Hodaei2017Nat,Chen2017Nat} and quantum regimes~\cite{Brody2012PRL,Lee2014PRX,Li2019NatCom,Kawabata2017PRL,Hamazaki2019PRL,Xiao2019PRL,Wu2019Science,Yamamoto2019PRL,Yamamoto2019PRL,Naghiloo2019NatPhys,Matsumoto2020PRL}. Non-Hermitian systems exhibit various exotic properties that cannot be found in Hermitian systems. For instance, the energy spectra of non-Hermitian Hamiltonians are normally complex, but it can be real when $\mathcal{PT}$ symmetry~\cite{Bender1998PRL,Bender2002PRL,Bender2007RPP} or pseudo-Hermiticity~\cite{Mostafazadeh2002JMP,Mostafazadeh2010IJMMP,Moiseyev2011Book,Zeng2020PRB1,Kawabata2020PRR,Zeng2021NJP} is imposed. On the other hand, one of the most extraordinary phenomena in non-Hermitian systems is the non-Hermitian skin effect (NHSE), where a large number of eigenstates are accumulated at the boundary of the system~\cite{Yao2018PRL1,Yao2018PRL2}. The existence of NHSE can modify the system's properties substantially and has attracted intensive attention in recent years~\cite{Alvarez2018PRB,Alvarez2018EPJ,Lee2019PRB,Zhou2019PRB,Kawabata2019PRX,Song2019PRL,Okuma2020PRB,Xiao2020NatPhys,Yoshida2020PRR,Longhi2019PRR,Yi2020PRL,Claes2021PRB,Haga2021PRL,Zeng2022PRA,Zeng2022PRB}. In topological systems, it has been shown that the band topology can be altered in a significant way and the conventional principle of bulk-boundary correspondence breaks down due to the NHSE~\cite{Yao2018PRL1,Yao2018PRL2,Kunst2018PRL,Jin2019PRB,Yokomizo2019PRL,Herviou2019PRA,Zeng2020PRB,Borgnia2020PRL,Yang2020PRL2,Zirnstein2021PRL,Zhang2023SciBu}. Furthermore, the NHSE has a significant impact on the Anderson localization phenomenon, where mobility edges can be induced and the eigenenergies of the localized and extended states are found to exhibit different topological structures~\cite{Shnerb1998PRL,Gong2018PRX,Jiang2019PRB,Zeng2020PRR,Liu2021PRB1,Liu2021PRB2}. The spectra of such systems are also sensitive to changes in boundary conditions~\cite{Xiong2018JPC} and can be applied in designing quantum sensors~\cite{Budich2020PRL,Koch2022PRR}. 

The origin of NHSE under open boundary condition (OBC) is closely connected with the point gap in the energy spectrum of the non-Hermitian system under perioidc boundary condition (PBC)~\cite{Okuma2020PRL,Zhang2020PRL}. Further investigation of the NHSE has revealed that by varying the strength and range of the asymmetric hopping, more exotic features can emerge. For example, by extending the range of asymmetric hopping beyond the nearest-neighboring sites, the NHSE edge will emerge~\cite{Zeng2022PRB2}. However, most studies so far mainly focus on the NHSE in systems with constant asymmetric hopping, which corresponds to a constant imaginary vector potential or gauge field. If the imaginary vector potential becomes spatially varied, what will happen to the properties of non-Hermitian systems remains elusive. Moreover, previous research work on non-Hermitian physics mainly relies on tight-binding models, which are suitable for narrow-band systems but fails to describe wide-band states. So, it will also be interesting to investigate the non-Hermitian continuous systems and check whether the energy spectrum and NHSE will behave differently. 

In this work, we answer the above questions by studying the one-dimensional (1D) system subjected to a linearly varying imaginary vector potential, which is described by the continuous Schr\"odinger equation. By analytically solving the equation, we find that the eigenenergy spectrum of the system is always real under OBC but forms a parabola in the complex energy plane under PBC. All the eigenstates exhibit a modulated Gaussian distribution and are pinned on the same position, which is determined by the imaginary vector potential and boundary conditions. These behaviors are in sharp contrast to the NHSE in systems with constant imaginary vector potential, where the eigenstates are exponentially localized at the boundaries under OBC or become extended under PBC. When the zero point of imaginary vector potential is localized at the boundaries, the eigenstates are pinned on the left or right end of the 1D system, forming a Gaussian type of NHSE under OBC. Even though the energy spectrum under PBC forms an open curve, the Gaussian NHSE still has a topological origin and is characterized by a nonzero winding number in the PBC spectrum. The energies interior to the parabolic curve are shown to sustain localized edge states under semi-infinite boundary condition. The corresponding tight-binding lattice model with linearly varying asymmetric hopping between the nearest-neighboring sites is also investigated, revealing similar properties except that the PBC spectrum form closed loops instead of open curves. Our work sheds light on the exotic properties of non-Hermitian systems with spatially varying imaginary vector potentials.

The remaining sections of this paper are organized as follows. In Sec.~\ref{Sec2}, we introduce the 1D system subjected to a linearly varying imaginary vector potential and analytically solve the continuous Schr\"odinger equation. In Sec.~\ref{Sec3}, we discuss the system under semi-infinite boundary conditions and examine the Gaussian NHSE under OBC. We then investigate the corresponding tight-binding lattice model in Sec.~\ref{Sec4}. Finally, we conclude with a summary in Sec.~\ref{Sec5}.

\section{Continuous model}\label{Sec2}
We consider the particle moving in a 1D space described by the following Hamiltonian
\begin{equation}\label{H}
	H = \frac{1}{2m} \left[ p+i\gamma (x-x_0) \right]^2,
\end{equation}
where $p=-i\hbar\partial/\partial x$ is the momentum operator and $i\gamma (x-x_0)$ is a linearly varying imaginary vector potential or gauge field. Here, $\gamma$ is a real number that characterizes the rate of change of the vector potential and $x_0$ represents the zero point of the potential. The continuous systems with constant imaginary vector potentials have been studied before in several different situations~\cite{Hatano1996PRL,Hatano1997PRB,Longhi2021PRB}. The model studied here, however, has a linearly varying vector potential $A=i \gamma (x-x_0)$. For simplicity, we set $\hbar=2m \equiv 1$. Then the scaled form of the Schr\"odinger equation becomes
\begin{equation}\label{SchEq1}
	-\left[ \frac{\partial}{\partial x} + \gamma (x-x_0) \right]^2 \psi (x)= E\psi(x),
\end{equation}  
with $E$ being the eigenenergy and $\psi(x)$ the corresponding wave function. Suppose that the particle is confined in the region $0 \leq x \leq L$, then we have open boundary condition as 
\begin{equation}
	\psi(0) = \psi(L) = 0,
\end{equation}
or periodic boundary condition as
\begin{equation}
	\psi(0) = \psi(L).
\end{equation}

To solve the continuous Schr\"odinger equation in Eq.~(\ref{SchEq1}), we take the following imaginary gauge transformation
\begin{equation}
	\psi(x)=\varphi(x) \exp \left[ -\frac{1}{2} \gamma (x-x_0)^2 \right],
\end{equation}
then the equation reduces to 
\begin{equation}\label{SchEq2}
	-\frac{\partial^2}{\partial x^2}\varphi(x)=E\varphi(x)
\end{equation}
with $\varphi(0)=\varphi(L)=0$ under OBC. Notice that the imaginary gauge transformation method introduced here can be generalized to systems subject to imaginary vector potentials of the type $A_m = i \gamma (x-x_0)^m$ with $m=0,1,2,\cdots$ (see the Appendix for details). The $m=0$ case corresponds to the system with a constant imaginary vector potential. Here we mainly focus on the linearly varying case with $m=1$, which corresponds to a constant imaginary magnetic field  under the Landau gauge: $\boldsymbol{B}=\bigtriangledown \times \boldsymbol{A}$ with $\boldsymbol{A}=\left[ i\gamma(x-x_0), 0 , 0 \right]$.  The imaginary magnetic fields have been utilized to explore the Lee-Yang zeros~\cite{Peng2015PRL} and $\mathcal{PT}$-symmetry breaking~\cite{Galda2018PRB} in spin systems. Here we will check the influences of imaginary magnetic field on the energy spectra and eigenstates of single particle systems.

\begin{figure}[t]
	\includegraphics[width=3.4in]{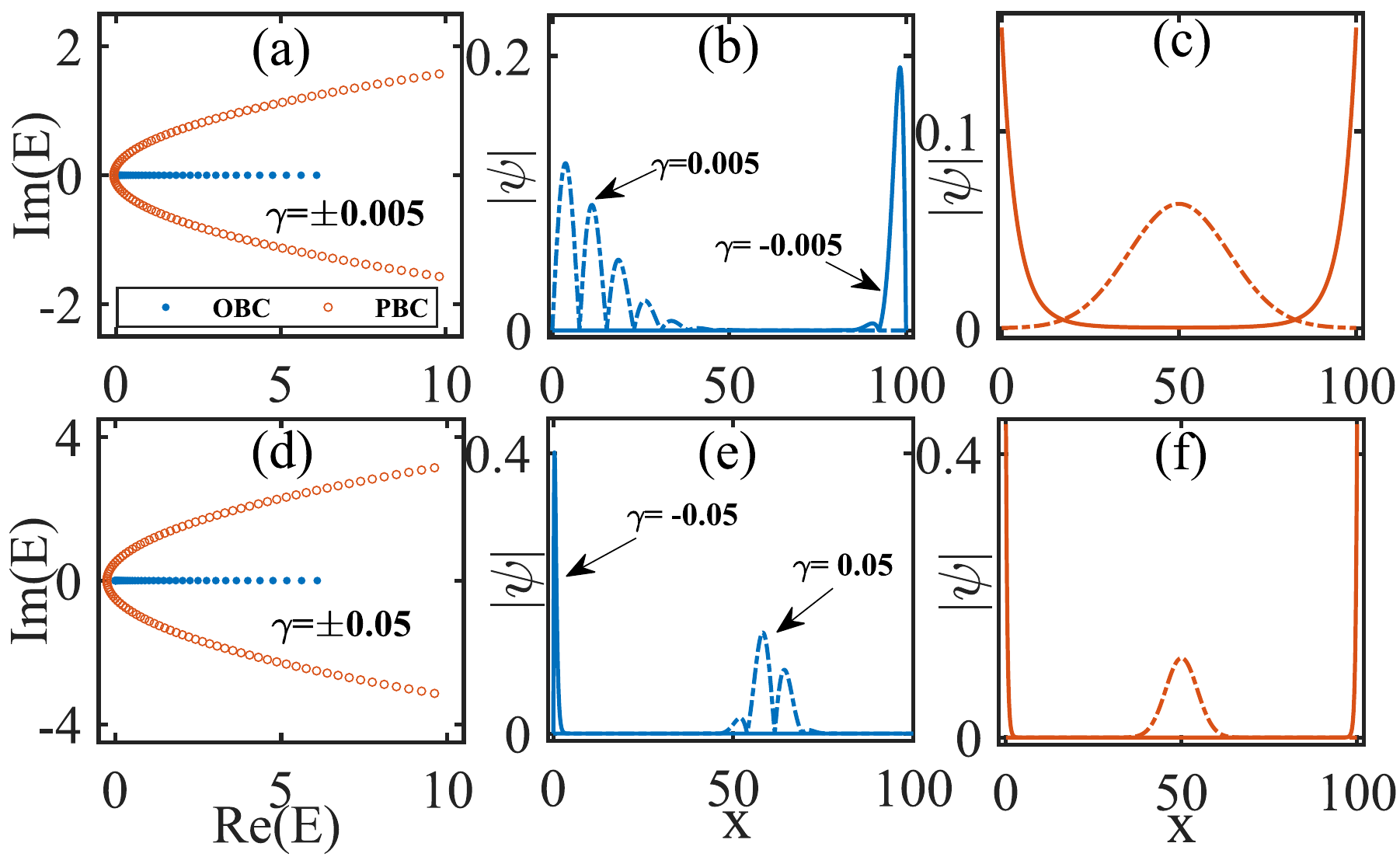}
	\caption{(Color online) (a) Eigenenergy of the non-Hermitian Hamiltonian in Eq.~(\ref{H}) with $\gamma=\pm 0.005$ under OBC (blue solid dots) and PBC (red empty dots). (b) and (c) shows the distribution of eigenstates corresponding to $E_n$ with $n=13$ for the system with $\gamma=0.005$ (dot-dashed line) and $-0.005$ (solid line) under OBC and PBC, respectively (The complex energies are sorted according to their real parts). Here we set $0\leq x \leq 100$ and $x_0=0$. (d)-(f) are the same as (a)-(c) but with $\gamma=\pm 0.05$ and $x_0=60$.}
	\label{fig1}
\end{figure}

After the transformation, we can see that Eq.~(\ref{SchEq2}) is the equation of a free particle confined in $0 \leq x \leq L$, which can be easily solved to obtain
\begin{equation}\label{E_OBC}
	\varphi_n^{OBC} (x)=\sqrt{\frac{2}{L}} \sin \left( \frac{n\pi x}{L} \right), \quad E_n^{OBC} = \frac{n^2 \pi^2}{L^2}
\end{equation}
with $n=1,2,\cdots$. So, the corresponding wave function for the original Schr\"odinger equation Eq.~(\ref{SchEq1}) is
\begin{equation}\label{Ev_OBC}
	\psi_n^{OBC} (x) = C_1 \exp \left[ -\frac{1}{2} \gamma (x-x_0)^2 \right] \sin \left( \frac{n\pi x}{L} \right),
\end{equation} 
where $C_1$ is for normalization. From this expression, we can see that all the wave functions exhibit a modulated Gaussian distribution and they are all pinned on $x=x_0$ when $\gamma>0$. If $\gamma<0$, then the states will be shifted to either the left or the right end depending on whether $x_0>L/2$ or not, see the distribution of eigenstates shown in Fig.~\ref{fig1}(b) and \ref{fig1}(e) for the systems with $x_0=0$ and $60$, respectively. If $x_0=L/2$, then the states will mainly distribute at the two ends. This is different with the system subjects to constant imaginary vector potential, where NHSE emerges under OBC and the states are exponentially distributed in the system. In addition, the eigenenergies are given by $E_n^{OBC}$, which means that the energy spectrum is real under OBC and is independent of the value of $\gamma$, as indicated by the blue dots in Fig.~\ref{fig1}(a) and \ref{fig1}(d).

Under PBC, on the other hand, the solution of Eq.~(\ref{SchEq2}) should satisfy the condition $\varphi(0)=\varphi(L)$, which leads to the following eigenstate
\begin{equation}
	\varphi_n^{PBC}(x) = \frac{1}{\sqrt{L}} \exp \left( i\frac{2\pi n x}{L} \right).
\end{equation}
and eigenenergy 
\begin{equation}\label{E_PBC}
	E_n^{PBC} = \left[ \frac{2n\pi}{L} + i \gamma \left(x_0-\frac{L}{2} \right) \right]^2
\end{equation}
with $n=0,\pm1,\pm2,\cdots$. This is also the eigenenergies of Hamiltonian $H$ under PBC, which is complex except for $x_0=L/2$. The corresponding eigenstate for Eq.~(\ref{SchEq1}) under PBC is 
\begin{equation}\label{Ev_PBC}
	\psi_n^{PBC}(x) = C_2 \exp \left[-\frac{1}{2} \gamma \left(x-\frac{L}{2} \right)^2 + i\frac{2\pi n x}{L} \right].
\end{equation}
So the eigenstates are also of Gaussian form and will always be pinned on the central position $x=L/2$ under PBC when $\gamma>0$, no matter where the zero point of the imaginary vector potential is. If $\gamma<0$, however, the states will be shifted to the position with $x=0(L)$ [see Figs.~\ref{fig1}(c) and \ref{fig1}(f)]. This is very different from the case under OBC, where the pinning position of the eigenstates is determined by the zero point and the sign of the imaginary vector potential. The behavior also differs significantly from the states in the system with constant imaginary vector potential, where the states are extended states that distribute over the whole system under PBC.

Let us further check the energy spectrum under different boundary conditions. From Eqs.~(\ref{E_PBC}) and (\ref{E_OBC}), we find that the OBC spectrum is real while the PBC spectrum forms a parabola in the complex energy plane, as shown by the blue solid dots and the red empty dots in Figs.~\ref{fig1}(a) and (d). Notice that the OBC spectrum is always located at the internal region of the open curve formed by the PBC spectrum with the same system parameters. The spectral properties are quite different from those of tight-binding lattice models, where the PBC spectra always form closed loops in the complex energy plane. The parabolic PBC spectrum has also been reported recently in continuous models with constant imaginary vector potential~\cite{Longhi2021PRB}.

\section{Edge modes in semi-infinite system}\label{Sec3}
\begin{figure}[t]
	\includegraphics[width=3.3in]{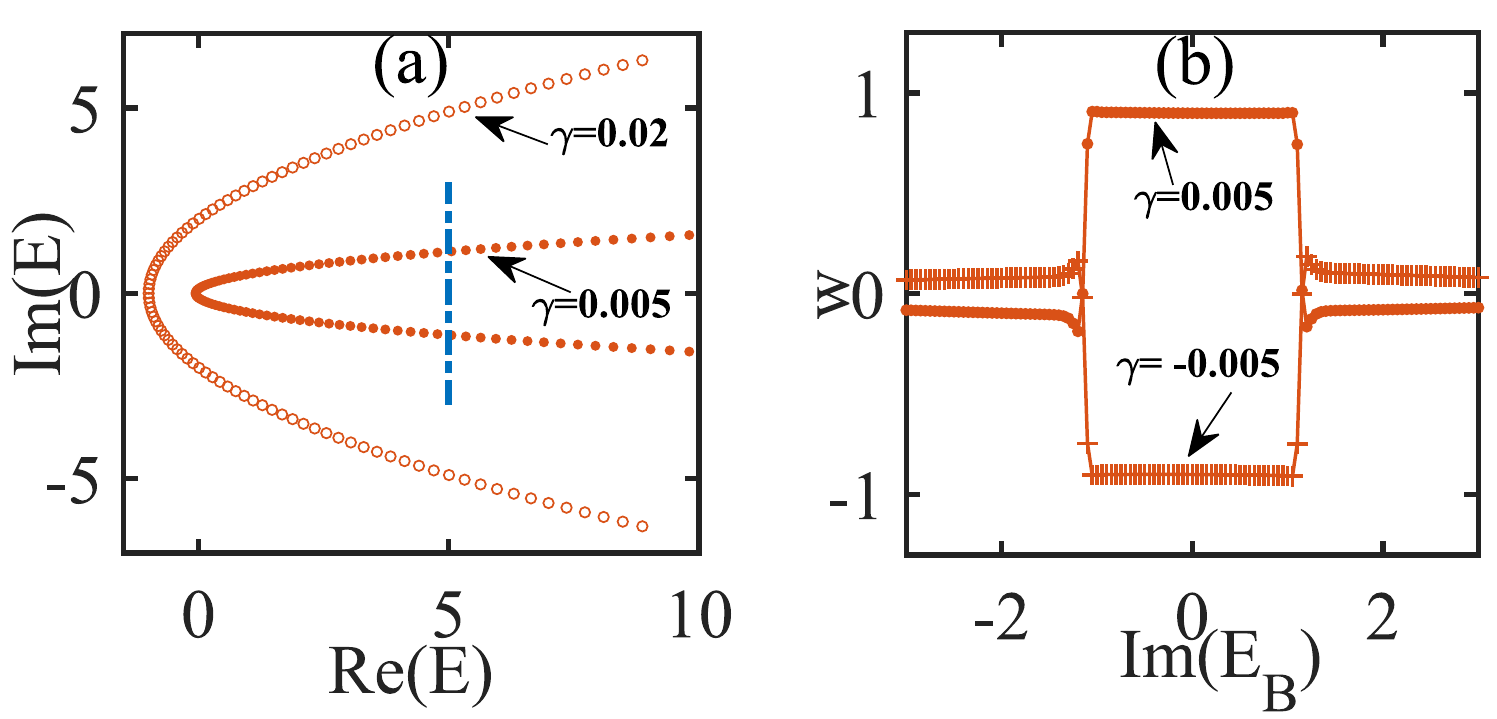}
	\caption{(Color online) (a) PBC spectrum for the system with $\gamma=0.02$ (red empty dots) and $\gamma=0.005$ (red solid dots). (b) Winding number as a function of the imaginary part of $E_B$ indicated by the blue dot-dashed line in (a). Here we set $0\leq x \leq 100$ and $x_0=0$.}
	\label{fig2}
\end{figure}

Next we consider the continuous model under the semi-infinite boundary condition (SIBC). To simplify the discussion, we will take $x_0=0$ in the following. From the above section, we know that under OBC, the eigenstates always exhibit a modulated Gaussian distribution and are localized at the left boundary when $\gamma>0$ but shifted to the right boundary when $\gamma<0$. Similar to the NHSE in systems with constant imaginary vector potential, here we can also explain the Gaussian type of NHSE as the edge modes of the system under SIBC and connect it with the nonzero winding number of the PBC spectrum. To prove this, here we focus on the case with $\gamma>0$ and set the SIBC as
\begin{equation}
	\psi(0)=0, \qquad \lim_{x\rightarrow +\infty} \psi(x) = 0.
\end{equation}
Suppose we choose $0<\gamma_1<\gamma_2$, and $f(x)$ as the eigenstate given in Eq.~(\ref{Ev_PBC}) satisfying the following Schr\"odinger equation under PBC
\begin{equation}\label{Hg1}
	H_{\gamma_1} f(x) = -\left[ \frac{\partial}{\partial x} + \gamma_1 x \right]^2 f(x)= E_B f(x),
\end{equation}
with $E_B$ being the corresponding eigenenergy. From $E_n^{PBC}$ in Eq.~(\ref{E_PBC}), we know that the PBC spectrum corresponding to $H_{\gamma_1}$ is interior to that of $H_{\gamma_2}$, as shown in Fig.~\ref{fig2}(a). 

Now we can prove that all the energies interior to the PBC spectrum of $H_{\gamma_2}$ are also eigenenergies of it under SIBC. To do so we can construct a wave function as 
\begin{equation}
	\psi_1(x) = f(x) e^{-\frac{1}{2}(\gamma_2-\gamma_1)x^2}
\end{equation} 
such that $\psi_1(x)$ satisfies the Schr\"odinger equation
\begin{equation}\label{Hg2}
	H_{\gamma_2} \psi_1(x) = -\left[ \frac{\partial}{\partial x} + \gamma_2 x \right]^2 \psi_1(x)= E_B \psi_1(x).
\end{equation}
This means that $E_B$ is also an eigenenergy of Hamiltonian $H_{\gamma_2}$. And the wave function $\psi_1(x)$ decays to zero as $x\rightarrow +\infty$. The other solution to the above second-order differential equation can be obtained by using the Liouville's formula, which gives 
\begin{equation}
	\psi_2(x) = \psi_1(x) \int dx \frac{e^{-\gamma_1 (x-x_0)^2}}{\psi_1^2(x)} = \psi_x e^{-(\gamma_1 L + i \frac{4\pi n}{L})x}.
\end{equation} 
$\psi_2(x)$ also decays to zero as $x\rightarrow +\infty$. So the general solution to the differential equation in Eq.~(\ref{Hg2}) is 
\begin{equation}
	\psi(x) = A \psi_1(x) + B \psi_2(x),
\end{equation}
where $A$ and $B$ are arbitrary constants. To satisfy the condition $\psi(0)=0$, we can set $A=-B=C$ and get 
\begin{equation}
	\psi(x) = C \left[ \psi_1(x) - \psi_2(x) \right].
\end{equation}
Apparently, the above $\psi_(x)$ satisfies the condition $lim_{x\rightarrow +\infty} \psi_(x) = 0$. Thus the wave function $\psi(x)$ is the eigenstate of $H_{\gamma_2}$ with the eigenergy $E_B$. For any energy interior to the parabola corresponding to $H_{\gamma_2}$ under PBC, they are also the eigenenergies of $H_{\gamma_2}$ under SIBC and the corresponding wave functions are the edge states localized at the left end of the semi-infinite system. For $\gamma<0$, we can also obtain a similar conclusion with the wave function localized at the right end instead. 

Similar to the topological origin of NHSE in systems with constant imaginary vector potential, we can also define a winding number for $E_B$ interior to the PBC spectrum of $H_\gamma$ as 
\begin{equation}
	w = \frac{1}{2\pi i} \int_{-\pi}^{\pi} dk \frac{d}{dk} \log \left[ E_\gamma^{pbc}(k) - E_B \right]
\end{equation}  
with $E_\gamma^{pbc}(k)$ given in Eq.~(\ref{E_PBC}) and $k=2n\pi/L$. $E_B$ is the base energy. When $\gamma>0$ and $E_B$ is interior to the parabola corresponding to the PBC spectrum, we have $w=1$; while if $\gamma<0$, we have $w=-1$, as shown in Fig.~\ref{fig2}(b). For the base energy $E_B$ indicated by the blue dot-dashed line in Fig.~\ref{fig2}(a), the winding number is close to $\pm 1$ when $E_B$ is interior to the PBC spectrum of $\gamma=\pm 0.005$. However, when $E_B$ moves out of the internal regime, the winding number drops sharply to almost 0. So even though the PBC spectrum forms a parabola, which is an open curve, the winding number can still be nonzero and quantized. This can be explained from Eq.~(\ref{E_PBC}), where we can find that as $|n|$ increases, the imaginary part grows slower than the real part, so we have
\begin{equation}
	\lim_{|n| \rightarrow +\infty} \frac{Im(E_n^{PBC})}{Re(E_n^{PBC})} = 0.
\end{equation}
Thus the PBC spectrum can be taken as a closed loop and we can obtain a quantized winding number. The skin modes are the edge modes corresponding to the nonzero winding number under PBC. So similar to the non-Hermitian systems with constant imaginary vector potential, the Gaussian NHSE here can also be taken as the edge states of the system under SIBC and is tightly connected to the point gap in the PBC spectrum.

\section{Tight-binding model}\label{Sec4}
\begin{figure}[t]
	\includegraphics[width=3.4in]{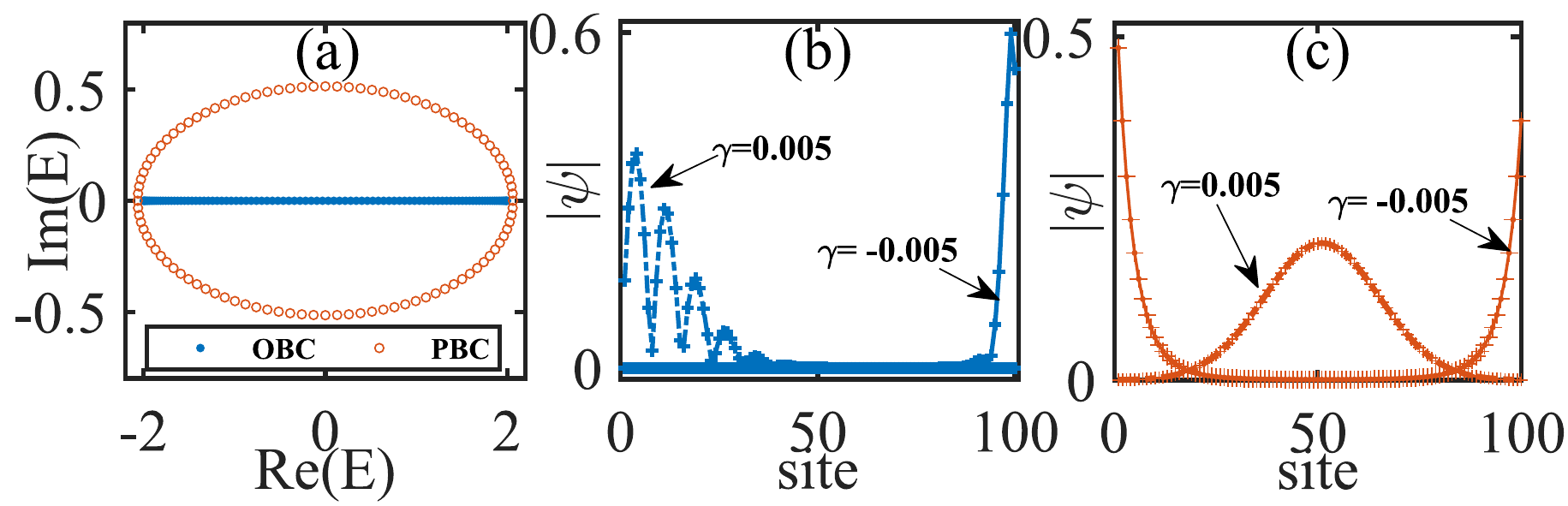}
	\caption{(Color online) (a) OBC (blue solid dots) and PBC (red empty dots) spectrum for the 1D tight-binding lattice with $\gamma=\pm 0.005$. (b) and (c) show the distribution of eigenstates under OBC and PBC. The lines represents the numeric results from diagonalization while the "+" signs represent the eigenstates obtained from the analytical expression. Here we set $N = 100$ and $j_0=0$.}
	\label{fig3}
\end{figure}

The continuous model discussed above can be transferred into a lattice model under the tight-binding approximation, which leads to the following model Hamiltonian
\begin{equation}
	H = \sum_j \left[ t e^{\gamma (j-j_0+\frac{1}{2})} c_j^\dagger c_{j+1} + t e^{-\gamma (j-j_0+\frac{1}{2})} c_{j+1}^\dagger c_j \right].
\end{equation}
Here we set the lattice constant to be 1. $c_j$ ($c_j^\dagger$) is the annihilation (creation) operator at the $j$th site. $t e^{\gamma (j-j_0+\frac{1}{2})}$ and $t e^{-\gamma (j-j_0+\frac{1}{2})}$ are the backward and forward hopping between the nearest-neighboring sites, respectively. $j_0$ is the referring site where the asymmetric hopping is zero. The total lattice number of the lattice is $N$. By diagonalizing the Hamiltonian, which can be represented as an $N \times N$ matrix, we can get the energy spectrum under OBC and PBC, as shown by the solid dots in Fig.~\ref{fig3}(a). The eigenstates of the tight-binding Hamiltonian can be obtained by rewriting the wave functions $\psi_n^{OBC}$ and $\psi_n^{PBC}$ shown in Eqs.~(\ref{Ev_OBC}) and (\ref{Ev_PBC}) in the continuous model and construct the eigenstate as follows
\begin{align}
	\psi_{n,j}^{OBC} &= \exp \left[-\frac{1}{2} \gamma (j-j_0)^2 +i \pi j \right] \sin \left( \frac{n \pi j}{N+1} \right); \\
	\psi_{n,j}^{PBC} &= \exp \left[ -\frac{1}{2} \gamma \left(j- \lceil \frac{N+1}{2} \rceil \right)^2 + i \frac{2\pi n j}{N} \right].
\end{align}
Here $x \leq \lceil x \rceil \leq (x+1)$ is the ceiling function. Then the wave function of lattice can be written as $\Psi_n^{OBC} =\sum_j \psi_{n,j}^{OBC}$ and $\Psi_n^{PBC} =\sum_j \psi_{n,j}^{PBC}$, respectively. The wave function satisfies the discrete Schr\"odinger equation
\begin{equation}
	E_n^t \psi_{n,j} = t e^{-\gamma (j-j_0+\frac{1}{2})} \psi_{n,j-1} + t e^{\gamma (j-j_0+\frac{1}{2})} \psi_{n,j+1}. 
\end{equation}
Solving the equation gives us the following eigenenergy spectrum
\begin{align}
	E_n^{t-OBC} &= 2 t \cos \frac{n\pi}{N+1},\\
	E_n^{t-PBC} &= 2 t \cos \left[ \frac{2\pi n}{N} + i \gamma \left(j_0 -\frac{N+1}{2} \right) \right],
\end{align}
where $n=1,2,\cdots,N$. So the spectrum under OBC is always real; while the spectrum under PBC is similar to the regular spectrum of the 1D tight-binding lattice but replaces the momentum $k$ by a complex one $k \rightarrow k + i \gamma \left(j_0 -\frac{N+1}{2} \right)$ with $k=2\pi n /N$. The energy spectrum using the above analytical expressions is totally consistent with the numerical results in Fig.~\ref{fig3}(a). Different from the continuous model where the spectrum under PBC forms a parabola, here the PBC spectrum of the tight-binding lattice forms a closed loop, which encloses the spectrum under OBC. For the tight-binding Hamiltonian matrix under OBC, we can similarly transform it into an Hermitian matrix through a similarity transformation as $h=S H S^{-1}$, where $S$ is a diagonal matrix $S=diag(e^{\gamma s(1)}, e^{\gamma s(2)}, \cdots, e^{\gamma s(N)})$ with $s(1)=1-j_0+\frac{1}{2}$ and $s(j)=s(j-1)+(j-1)-j_0+\frac{1}{2}$. The transformed matrix corresponds to the tight-binding lattice with constant hopping $t$:
\begin{equation}
	h = \sum_j t c_j^\dagger c_{j+1} + t c_{j+1}^\dagger c_j.
\end{equation}
So the OBC spectrum of the lattice is always real. 

In Figs.~\ref{fig3}(b) and (c) we gives the numerical results of the distributions of wave functions of the system under OBC and PBC in the lattice with $x_0=0$. They can be fully fitted by our analytical expressions $\psi_{n,j}^{OBC}$ and $\psi_{n,j}^{PBC}$. When under OBC, we can see that the eigenstates are localized at the left (right) end of the 1D lattice if $\gamma>0$ ($\gamma<0$). This is the Gaussian NHSE, and its origin roots in the point gap in the PBC spectrum as shown in Fig.~\ref{fig3}(a). These features are the same as the continuous model, except that the PBC spectrum forms closed loops instead of open curves.

\section{Summary}\label{Sec5}
In this work, we study the one-dimensional systems with a linearly varying imaginary vector potential described by the single-particle continuous Schr\"odinger equation, which is analytically solved. Our findings show that the eigenenergies are real under OBC but form a parabola in the complex energy plane under PBC. Additionally, we discover that all the eigenstates exhibit a modulated Gaussian distribution and are pinned at the same position, which is determined by the imaginary vector potential and boundary conditions. This behavior is in stark contrast to systems with constant vector potential. We further demonstrate that the Gaussian type of NHSE under OBC is associated with the nonzero winding number in the PBC spectrum. Finally, we extend our study to a tight-binding lattice model, where similar phenomena are found except that the PBC spectrum forms closed loops instead of open curves. Overall, our work expands the understanding of non-Hermitian physics and opens new avenues for studying quantum systems with spatially varying imaginary vector potentials.

\section*{Appendix}\label{Append}
In this Appendix, we give the gauge transformation for solving the Schr\"odinger equation with a general imaginary vector potential of the type
\begin{equation}
	A_m = i \gamma (x-x_0)^m, \quad m=0,1,2,\cdots
\end{equation}
Notice that the $m=0$ and $m=1$ case corresponds to the system with constant or linearly varying imaginary vector potential, respectively. For the general cases, the Hamiltonian of the system is 
\begin{equation}
	H  = \frac{1}{2m} \left[ p+ i \gamma (x-x_0)^m \right]^2.
\end{equation}
After setting $\hbar=2m \equiv 1$, we can obtain the following Schr\"odinger equation as
\begin{equation}\label{SchEq1}
	-\left[ \frac{\partial}{\partial x} + \gamma (x-x_0) \right]^2 \psi (x)= E\psi(x),
\end{equation}  
where $E$ is the eigenenergy and $\psi(x)$ is the corresponding wave function. To solve this equation, we can take the following gauge transformation
\begin{equation}
	\psi(x) = \varphi (x) \exp \left[ -\frac{1}{m+1} \gamma (x-x_0)^m \right].
\end{equation}
Then the equation will be reduced to
\begin{equation}
	-\frac{\partial^2}{\partial x^2}\varphi(x)=E\varphi(x),
\end{equation}
which is the Schr\"odinger equation for a free particle and can be solved easily combining with the appropriate boundary conditions. The imaginary gauge transformation introduced here can be used to solve this type of Schr\"odinger equation.

\begin{acknowledgments}
This work is supported by NSFC (Grant No. 12204326), R\&D Program of Beijing Municipal Education Commission (Grant No. KM202210028017) and Open Research Fund Program of the State Key Laboratory of Low-Dimensional Quantum Physics (Grant No. KF202109). R.-L is supported by the NSFC under Grant No. 11874234 and the National Key Research and Development Program of China (Grant No. 2018YFA0306504)
\end{acknowledgments}

\end{document}